\documentclass[conference]{IEEEtran}
\usepackage[latin9]{inputenc}
\usepackage{amsmath}
\usepackage{graphicx}
\usepackage{epstopdf} 
\usepackage{amssymb}  
\usepackage{xcolor}   
\begin{document}
%
\title{
\vspace{-1.3cm}
{\footnotesize \textnormal{\textit{
IEEE International Workshop on Information Forensics and Security, December 9-12, 2019, Delft, The Netherlands\vspace{+0.7cm}\\}}}
Pooled Steganalysis in JPEG: \\how to deal with the spreading strategy?}

\author{\IEEEauthorblockN{Ahmad ZAKARIA}
\IEEEauthorblockA{LIRMM, Univ Montpellier,  \\ CNRS, \\ France. \\
Email: ahmad.zakaria@lirmm.fr}
\and
\IEEEauthorblockN{Marc CHAUMONT}
\IEEEauthorblockA{LIRMM, Univ Montpellier, \\ CNRS, Univ N\^imes,\\ France. \\
Email:  marc.chaumont@lirmm.fr}
\and
\IEEEauthorblockN{G\'erard SUBSOL}
\IEEEauthorblockA{LIRMM, Univ Montpellier,  \\ CNRS, \\ France. \\
Email:  gerard.subsol@lirmm.fr}}



\maketitle

\begin{figure}[b]
\vspace{-0.3cm}
\parbox{\hsize}{\em
}\end{figure}

\begin{abstract}
In image pooled steganalysis, a steganalyst, Eve, aims to detect if {\it a set} of images sent by a steganographer, Alice, to a receiver, Bob, contains a hidden message. We can reasonably assess that the steganalyst does not know the strategy used to spread the payload across images. To the best of our knowledge, in this case, the most appropriate solution for pooled steganalysis is to use a Single-Image Detector (SID) to estimate/quantify if an image is cover or stego, and to average the scores obtained on the {\bf {\it set}} of images. 

In such a scenario, where Eve does not know the spreading strategies, we experimentally show that if Eve can {\it discriminate} among few well-known spreading strategies, she can improve her steganalysis performances compared to a simple averaging or maximum pooled approach. Our {\it discriminative} approach allows obtaining steganalysis efficiencies comparable to those obtained by a clairvoyant, Eve, who knows the Alice spreading strategy. Another interesting observation is that DeLS spreading strategy behaves really better than all the other spreading strategies.

Those observations results in the experimentation with six different spreading strategies made on Jpeg images with J-UNIWARD, a state-of-the-art Single-Image-Detector, and a {\it discriminative} architecture that is invariant to the individual payload in each image, invariant to the size of the analyzed set of images, and build on a binary detector (for the pooling) that is able to deal with various spreading strategies. 
\end{abstract}


%
\IEEEpeerreviewmaketitle

\section{Introduction}
\label{sec:intro}


\textit{Steganography} consists in altering a digital object (called \textit{cover}), in an innocuously looking way, to hide (or embed) a message which allows secret message communication. The resulting altered object is named \textit{stego}. The science of {\it detection} of the presence of this hidden data, given an innocuous object, is called \textit{steganalysis}. In this paper, we focus on the steganalysis of digital images, which are the most studied cover objects. More precisely, we use images coded in JPEG format which is the most common one nowadays.

Steganography traditionally focused on embedding a message in one image at a time, but it is much more realistic for the steganographer to hide the message by {\it spreading} it over multiple images. This {\it spreading} is called \textit{batch steganography} and is to be opposed to the \textit{pooled steganalysis} where a {\it set} of images are analyzed in order to gather a set of clues, and thus to conclude to the presence/absence of a hidden message in the {\it set of images}. Batch steganography and pooled steganalysis topics were introduced in \cite{ker2006batch} and became one of the most challenging open problems in the field these last years \cite{ker2013moving}. 

\begin{figure}
\includegraphics[width=0.48\textwidth]{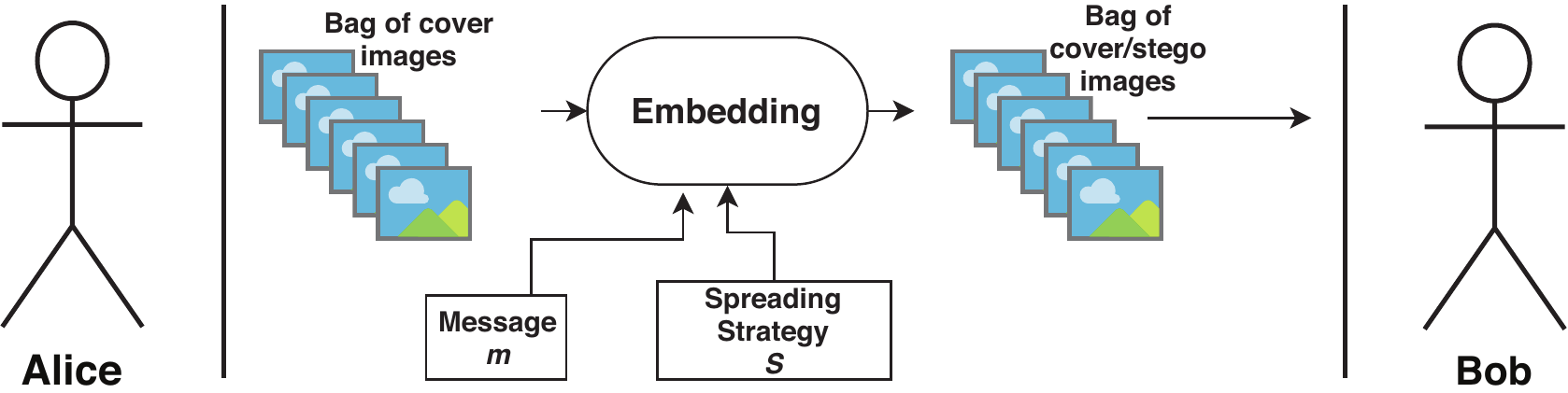}
\vspace{-0.1cm}
\caption{A scheme that illustrates how the steganographer, Alice, spreads a message ${\bf m}$ in multiple covers using a strategy $s \in \mathcal{S}$.}
\label{fig:IntroBatch}
\end{figure}

As presented in Figure \ref{fig:IntroBatch}, {\it batch steganography} consists in embedding a message ${\bf m} \in \{0,1\}^{|m|}$, of length $|m|$, in a bag of cover images by using a spreading strategy, $s\in\mathcal{S}$, from a set of possible spreading strategies $\mathcal{S}$. 

\begin{figure}
\vspace{-0.2cm}
\includegraphics[width=0.48\textwidth]{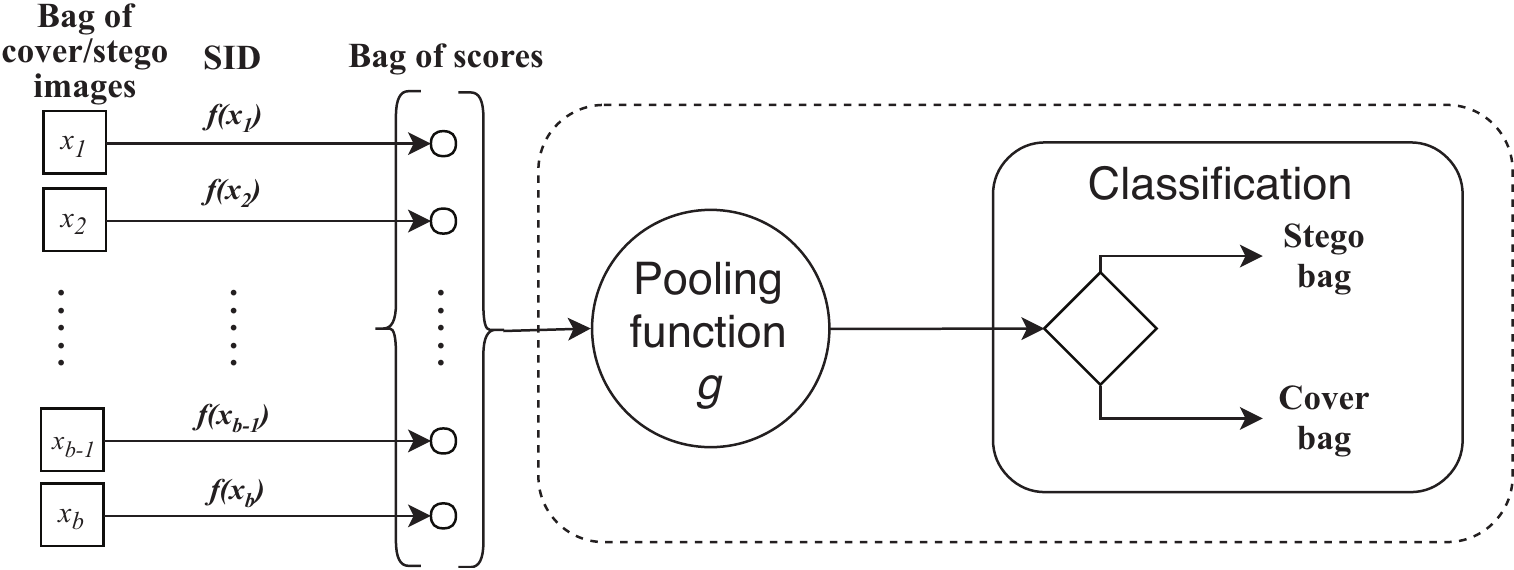}
\vspace{-0.1cm}
\caption{A scheme that illustrates how the steganalyst, Eve, uses a pooling function $g$ to aggregate evidence from multiple images in order to make a final decision about the presence/asbcence of a hidden message.}
\vspace{-0.1cm}
\label{fig:IntroPooled}
\end{figure}

As presented in Figure \ref{fig:IntroPooled}, modern {\it pooled steganalysis} consists to take a bag $B$ of $b$ images $\{x_1, ..., x_b\}$, which may be cover or stego, applies a Single Image Detector (SID), denoted by the function $f$, which can be based either on a binary or a quantitative algorithm, in order to get the set of SID scores $\{f(x_1), ..., f(x_b)\}$. Then the steganalyst aggregate these scores by using a pooling function $g:\mathbb{R}^b\mapsto \mathbb{R}$ in order to get a single output which allows to classify the bag as {\it cover} or {\it stego}. 


The first attempts in order to {\it experimentally} evaluate batch steganography / pooled steganalysis has started in 2011 and 2012 \cite{ker2011new, ker2012batch}, with the paradigm of {\it multiple users} (multiple actors), and the research of outliers. Instead of making an individual and independent binary decision on which actor is guilty, the proposed algorithms rank the actors according to their guiltiness. Nevertheless, those algorithms does not work in a {\it single actor} case, which is a more general, and thus more interesting working case.

Since 2015, three papers address the problem of only one actor and a pooling defined by the aggregation of clues (SID scores) computed on each image individually \cite{cogranne2015sequential}, \cite{pevny2015optimizing}, and \cite{cogranne2017practical}. 

In the article \cite{cogranne2015sequential}, Cogranne deals with the case where {\it Eve \underline{does not know} the spreading strategy} used by Alice. In that case, Cogranne shows that the best pooling strategy consists of {\it averaging} the individual scores. In the article \cite{pevny2015optimizing}, Pevny and Nikolaev deal with the case where {\it Eve \underline{does know} the spreading strategy} used by Alice. In that case, Pevny and Nikolaev observe that the knowledge of the strategy allows improving the steganalysis results. In the article \cite{cogranne2017practical}, Cogranne also shows that if {\it Eve \underline{does know} the spreading strategy} used by Alice, it allows to better aggregate the individual SID scores, and thus to obtain better steganalysis results. The tendency shared by those papers is that the optimal pooling function $g$ applied on the SID scores depends on the steganographer's strategy used to spread the messages among multiple objects (this was also expressed in the seminal paper \cite{ker2006batch} in the case of non-adaptive embedding).

In our paper, we study the scenario where {\it Eve \underline{does not know}} the spreading strategy used by Alice, and we propose to evaluate Eve's capability to obtain better results than those she would obtain by averaging the SID scores. To reach that goal, first, Eve {\it learns} well-known spreading strategies, and secondly, during the test time, we test Eve's ability to {\it discriminate} between a {\it cover} bag and a {\it stego} bag. During the test, Eve applies a weighted sum during pooling instead of an average. Our approach does not guess the spreading strategy used by Alice 
even if, thanks to the learning process, Eve is able to better {\it discriminate} between {\it cover} bag and {\it stego} bag, by distinguishing better the various statistics associated to each strategy. The reader must understand that compared to the previous papers and especially to \cite{pevny2015optimizing}, the question which is raised in our paper is first, about the ability for Eve to use some knowledge about the spreading strategies (without knowing exactly which one is used by Alice), and second, to do it in a more realist scenario (no knowledge about the individual payload sizes, no knowledge about the spreading strategy, and eventually no knowledge of the bag size).

In Section \ref{sec:state_of_the_art} we briefly describe a state-of-the-art list of spreading strategies and a we present a general pooled steganalysis architecture. In Section \ref{sec:proposition} we discuss more in detail the question raised in our paper. Finally, we present in Section \ref{sec:Results} the results and discuss them. We conclude and give some perspectives in Section \ref{sec:conclusion_and_perspectives}.

\section{Batch spreading strategies and a general pooled steganalysis architecture}
\label{sec:state_of_the_art}


\subsection{Batch spreading strategies ($\mathcal{S}$)} \label{subsec:practical batch strategies}

Note that the spreading in a bag is done at a given payload size which is expressed in bit per total coefficients (bptc). The bptc is thus the size of the message in bits divided per the total number of pixels (i.e. the ACs and DCs coefficients) of all the images in the bag. We have compiled the batch spreading strategies proposed in \cite{pevny2015optimizing, cogranne2017practical,ker2014steganographer} in the following list:
\begin{enumerate}

\item \textbf{Greedy strategy:} the steganographer embeds the message into as few covers as possible. The steganographer randomly chooses a cover and embeds part of his message up to 1 bit per coefficients (bpc)\footnote{The choice of 1 bpc for the maximum payload makes the construction of the bag practically easier.}, because the intrinsic security of the image is not taken into account in the greedy strategy. Eve repeats the embedding process with another randomly chosen cover until the whole message bits are embedded.

\item \textbf{Linear strategy:} the steganographer distributes the message evenly across all available covers, so the payload size, in bits, for each image is equal to the message length divided by $b$.

\item \textbf{$Uses$-$\beta$ strategy:} the steganographer distributes the message evenly across a fraction $\beta$ of available cover images. This strategy is equivalent to the greedy one for $\beta =\alpha$, and to the linear one for $\beta = 1$.

\item \textbf{Image Merging Sender (IMS) strategy:} the steganographer generates a unique image from all the $b$ images from a bag $B$, and lets the embedding algorithm spread the payload all over this ``big'' image. More precisely a cost value is computed for each DCT coefficient of the image (the adaptive embedding algorithm define the way the cost map is computed), and then the embedding is obtained with STC \cite{Filler2011STC} or the simulator \cite{fridrich2006minimizing}.

\item \textbf{Detectability Limited Sender (DeLS) strategy:} the steganographer adopts a cover model and spreads payload over $b$ images that communicates the required payload, so that each image from the bag contributes with the same value as the Kullback-Leibler (KL) divergence (deflection coefficient) based on MiPOD \cite{sedighi2016content} cover model\footnote{The deflection coefficient is computed on dequantized rounded images.}. 

\item \textbf{Distortion Limited Sender (DiLS) strategy:} The steganographer spreads payload over images so that each image from the bag contributes with the same value of distortion.

\end{enumerate}

\subsection{General pooled steganalysis architecture}
\label{sec:general_architecture}

\begin{figure} 
\includegraphics[width=0.48\textwidth]{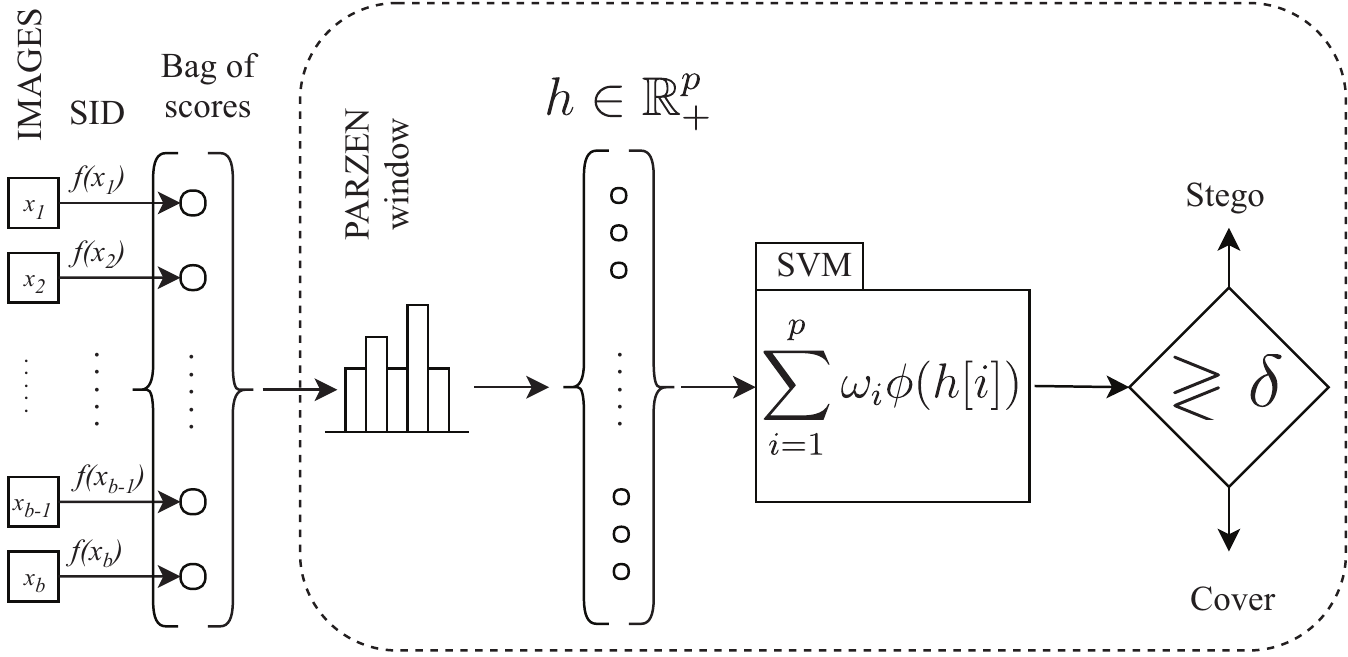}
\caption{The general pooled steganalysis architecture from \cite{pevny2015optimizing}. $\phi$ is the re-description transformation function. $\delta$ is a threshold.}
\label{fig:generalArchitecture}
\end{figure}

In this section we recall the general pooled steganalysis architecture that were proposed in \cite{pevny2015optimizing} for evaluating a given pooled steganalysis facing a given spreading strategy. Figure \ref{fig:generalArchitecture} resumes the different steps applied by Eve when she analyzes a bag of images. 

In the operational phase (i.e. when the general architecture is deployed), the pooled steganalysis algorithm takes as input a bag made of $b$ images $\{x_1, ..., x_b\}$ which may be cover or stego, and computes for each image the SID score, which is a real positive number, through the $f$ function (see Section \ref{sec:intro} and \ref{sec:proposition}). It thus gives a bag of real numbers $\{f(x_1), ..., f(x_b)\}$. From this bag made of $b$ values, a {\it Parzen window} (detailed below) is computed and lead to a histogram, noted ${\bf h}$, and made of $p$ bins. 
Finally, the pooling function aggregates the $p$ values of the histogram. 
The resulting weighted sum is then compared to a threshold, noted $\delta$ in order to decide if the bag is cover or stego.

Let us comment the Parzen window which is the most important ingredient of the architecture proposed in \cite{pevny2015optimizing}. The bag of SID scores, i.e. the vector ${\bf z}=\{f(x_1), ..., f(x_b)\}$, is transferred into a histogram representation thanks to the estimation by Parzen window. Given the Gaussian kernel function $k : \mathbb{R} \times \mathbb{R} \mapsto \mathbb{R}$ with $k(x, y) = exp(-\gamma||x-y||^2)$), the Parzen window computation is such that for a bag ${\bf z}$, the resulting histogram is: 

\begin{equation}\label{eq:histogram}
{\footnotesize {\bf h} =  \Bigg[ \frac{1}{b}\sum_{f(x_i)\in z}k(f(x_i),c_1),\ ...\ ,\ \frac{1}{b}\sum_{f(x_i)\in z}k(f(x_i),c_p) \Bigg] }
\end{equation}
with $\{c_i\}_{i=1}^p$ a set of equally spaced real positive values belonging to the range $min_{x\in\mathcal{X}} f(x)$ and  $max_{x\in\mathcal{X}} f(x)$, with $\mathcal{X}$ the images learning set. Each bin of the histogram ${\bf h}$, from Equation \ref{eq:histogram}, is the result of the cumulative Gaussian distance between each component of ${\bf z}$ and a scalar from the set of predefined centers $\{ c_i\}_{i=1}^p$.

Note that the histogram representation, ${\bf h}$, is of finite dimension $p$, whatever the dimension $b$ of the bag, and that this representation is invariant to the sequential order in the bag. 

Once the Parzen window is applied, the vector ${\bf h}$ of fixed dimension $p$ is given to an SVM which pools the vector component in the re-description space for classification:
\begin{equation}\label{fct:redescriptionSpace}
\sum_{i=1}^{p}\omega_{i}\ \phi(h[i]),
\end{equation}
with $\phi$ the function redefining the feature space. Note that $\phi$ is, in practice, never computed because of the "kernel trick". We can see when looking to Equation \ref{fct:redescriptionSpace}, that the pooling function is a weighted sum where the weights $\omega_{i}$ are learnt during the SVM training. It is clearly more subtle to pool the set of features $\{\phi(h[i])\}_{i=1}^p$ of the bag than using the straightforward average or maximum.

\section{Technical details}
\label{sec:proposition}

\subsection{Assumptions and limits of the general pooled steganalysis architecture}

In order to construct a general pooled steganalysis architecture, we should first choose a generic SID. It could be a quantitative detector such as \cite{kodovsky2013quantitative,pevny2015optimizing,zakaria2018quantitative,chen2018deep}, or a detector outputting a score such as \cite{cogranne2015modeling}. In this paper, we will use the quantitative detector described in \cite{kodovsky2013quantitative}.

To be the more realist possible, this SID should be invariant to image size (see some preliminary work in \cite{tsang2018steganalyzing}) for a given detectability and should be robust to cover-source-mismatch (CSM) and to the stego-mismatch. The CSM problem is described in \cite{cancelli2008comparative}; a holistic solution is proposed in \cite{lubenko2012going}, and an atomistic solution proposed in \cite{pasquet2014steganalysis}. An example of work related to the stego-mismatch can be found in \cite{Yousfi2019_Alaska}. In this paper, we will assume that the size of the images is fixed and that there is no cover-source mismatch, neither stego-mismatch. 

A general pooled steganalysis architecture must also be able to process a bag of any number of images. This is the case of the algorithms proposed in \cite{cogranne2015sequential} or \cite{pevny2015optimizing} even if in those papers, experiments are performed by using bags of only one size. On the contrary, this is not possible in the algorithm described in \cite{cogranne2017practical} where the hypothesis is that the steganalyst knows the number of images in the bag to analyze. In this paper, we propose an architecture dealing with any number of images in the bag, but in the experiments, we have trained our model on bags of fixed sizes. We postpone the experiment with bags of various size for future work.

Lastly, such architecture must be able to process a bag of any payload size, which is not done in any of the papers \cite{cogranne2015sequential, pevny2015optimizing, cogranne2017practical}. In our paper, we make the same assumption as \cite{cogranne2017practical} where the steganalyst learns at a fixed mean payload for each bag, and we postpone the experiments with any payload size in the bag for future work.

\subsection{Single Image Detector (SID)}

We choose as Single Image Detector (SID), which is referred in this paper as the function $f$, the feature-based Quantitative steganalysis algorithm proposed in \cite{kodovsky2013quantitative}, applied on the 17,000-dimensional JPEG domain Rich Model --- the Gabor features residuals (GFR) \cite{song2015steganalysis}. The Quantitative steganalysis algorithm is a machine learning regression framework that assembles, via the process of gradient boosting, a large number of simpler base learners built on random subspaces of the original high-dimensional feature space.

\subsection{Pooling functions}
\label{sec:general_architecture}

As explained in Section \ref{fig:generalArchitecture}, the general pooled steganalysis architecture provided in \cite{pevny2015optimizing} produces a feature space which is represented by a Parzen histogram, and proposes to pool the values of this histogram thanks to a linear SVM classifier. In the paper \cite{pevny2015optimizing}, only {\it one pooling} function (the SVM) is learnt for {\it one spreading strategy} since the study was on the comparison with the historical average and maximum pooling functions, depending on the embedded payload size in a bag. Additionally, the experiment done in \cite{pevny2015optimizing} only uses old spreading strategies (greedy and linear; see Section \ref{subsec:practical batch strategies}).

In our study, we look at the behaviour of the architecture when it has learned to recognize {\it various spreading strategy}. Our approach is thus a pooling function which is able to face multiple spreading strategies. The experiment objective is to show that even if Eve does not have any information on the spreading strategy used by Alice, she can obtain better steganalysis results than when using a simple average or maximum, and she can be close to the results that she would obtain if she was clairvoyant i.e. if she knows the spreading strategy. The difference compared with the paper \cite{pevny2015optimizing} is in the addressed question, and our paper is thus in the natural continuity of the three state-of-the-art papers \cite{pevny2015optimizing, cogranne2017practical,ker2014steganographer}. The experiments are in agreement with various spreading strategies (the modern and recent ones) and with state-of-the-art two-step machine learning in order to build the Rich Model and the SID.

In order to study the efficiency of our {\it discriminative} pooling function in order to {\it discriminate} $s$ among a set of strategies $\mathcal{S}$, we defined various pooling functions:
\begin{enumerate}
\item \textbf{$g_{disc}$:} This function is our {\it discriminative} pooling function and we train it on the Parzen histograms of all the strategies from $\mathcal{S}$. During the test, only one spreading strategy $s$ will be tested at a time. $g_{disc}$ is obtained through the learning of the various patterns from all the strategies, and the minimization of the classification error between a cover bag and a stego bag (whatever the spreading strategy) during the SVM learning.
Thanks to the Parzen representation, the SVM re-description space, and the weighted sum, the general architecture learns with different spreading strategies, and it should be able to classify better than applying an average or a maximum. 
\item \textbf{$g_{clair}$:} This function is the {\it clairvoyant} one. The training and the test are done with the knowledge of the used spreading strategy. The pooling function is obtained thanks to the use of an SVM, similarly as the $g_{disc}$ function.
\item \textbf{$g_{max}$:} This function is only a maximum applied on a Parzen histogram. The threshold $\tau_{max}$ is obtained by minimizing, on all the strategies, the total classification error probability under equal priors $P_e=\frac{1}{2}(P_{fa} + P_{md})$, where $P_{fa}$ and $P_{md}$ are the false-alarm and missed-detection probabilities. 
\item \textbf{$g_{mean}$:} This function is only an average applied on a Parzen histogram. The threshold $\tau_{mean}$ is obtained by minimizing, on all the strategies, the total classification error probability under equal priors $P_e=\frac{1}{2}(P_{fa} + P_{md})$, where $P_{fa}$ and $P_{md}$ are the false-alarm and missed-detection probabilities. 
\end{enumerate}


\section{Experimental evaluation}
\label{sec:Exp_Eval}

In this section, we compare $g_{disc}$ to $g_{mean}$, $g_{max}$ and $g_{clair}$. 



\subsection{Data preparation}\label{sec:Data_preparation}

Our image database is built from the BOSSbase 1.01 \cite{bas2011break}. We convert those 10 000 512 $\times$ 512 grey-scale {\bf spatial} images into JPEG images, using the MATLAB's command \textit{imwrite}, with quality factors 75. 

\begin{figure}
\includegraphics[width=0.48\textwidth]{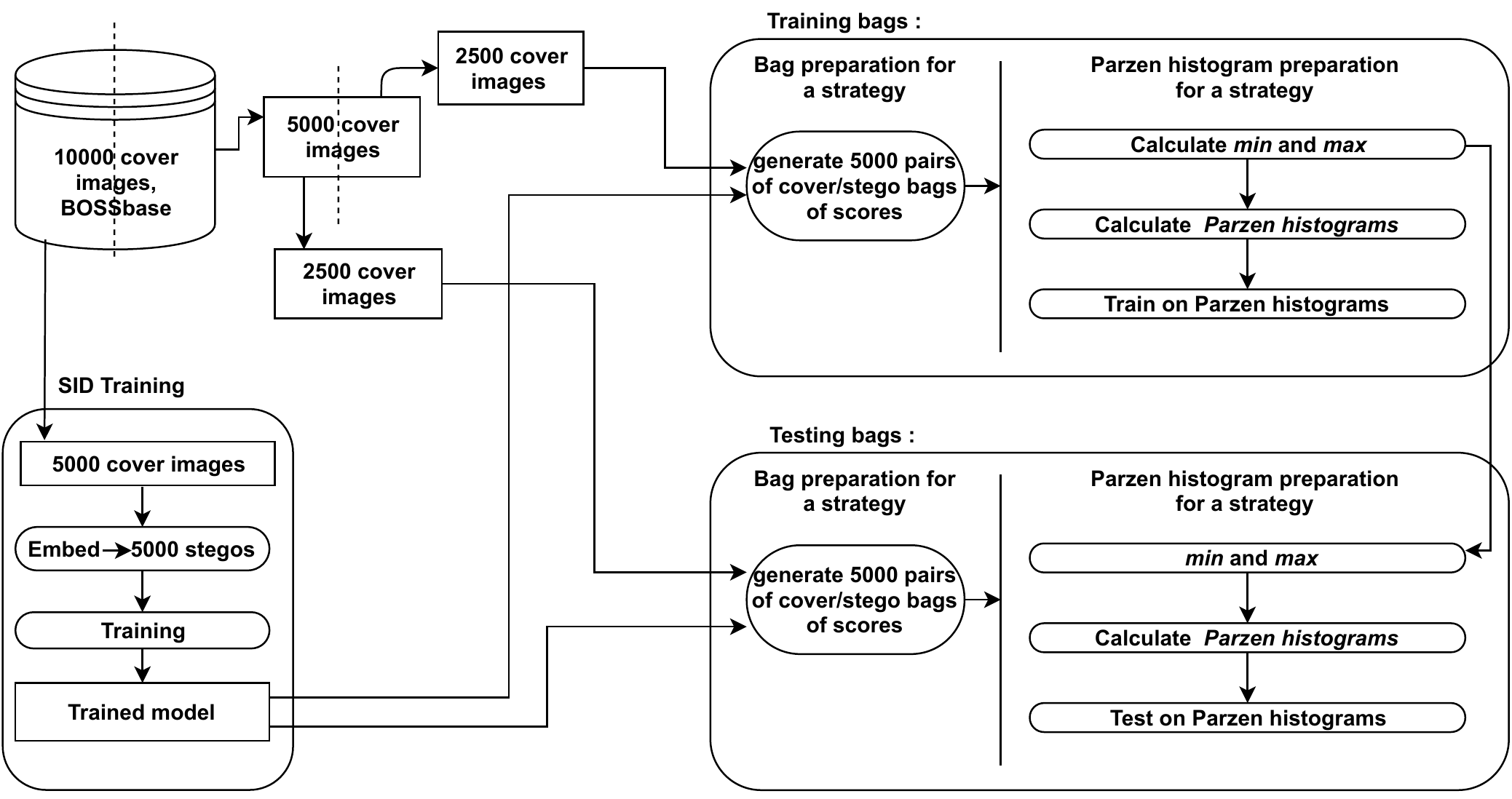}
\caption{Protocol for the data-sets creation, the learning and the test.}
\label{fig:splitting_data}
\vspace{-0.4cm}
\end{figure}

The 10 000 cover JPEG images are split in two equal sizes sets as shown in Figure \ref{fig:splitting_data}. The first set (5 000 cover images) is used for the learning of the quantitative SID. The second set (5 000 cover images) is used to create bags, learn the pooling functions, and test the various pooling.

Note that each time an embedding in an image is done, it is done with the J-UNIWARD scheme. 

Also note that each time a SID is used (in the first or second test, during the learning and also during the test, on a cover or on a stego whatever the payload size), for a given input image, a feature vector {\it Gabor Features Residuals} (GFR) \cite{song2015steganalysis} of dimension 17 000 is first extracted. This feature vector is then cleaned from NaN values (it occurs when the feature values are constant over images) and from constant values, to obtain a 16 750-dimensional feature vectors. Finally, we normalize this vector using the algorithm proposed in \cite{boroumand2017nonlinear}. 

\subsection{SID learning}
\label{sec:SIDlearning}

In order to train the quantitative Single Image Detector, we generate stegos such that there are 5 000 feature vectors for cover images and 5 000 feature vectors for {\it each} payload size whose range is fixed to \{0.1, 0.2, 0.3, 0.4, 0.5, 0.6, 0.7, 0.8, 0.9, 1\} bpc\footnote{We adapted the J-UNIWARD algorithm to insert an amount of data measured in bpc instead of bpnzAC.} which gives a total of 55 000 features vectors. Indeed, while training the quantitative SID, to avoid any bias in the results, we use 5 000 covers and 50 000 stegos such that the cover and each one of the ten different payload sizes are equally distributed. This way the resulted SID model will guarantee a fair scoring between each payload (payload 0 for covers). 

\subsection{Bags preparation}
\label{sec:bagspreparation}

The remaining 5 000 cover images are used for batch steganography i.e. bags preparation. More precisely 2~500 cover images are used for the learning, and the remaining 2~500 cover images will be used for the test. 

Given a bag size $b \in \mathcal{B}=\{2,$ $4,$ $6,$ $10,$ $20,$ $50,$ $100,$ $200\}$, and the set of spreading strategies $IMS$, $DeLS$, $DiLS$, $Greedy$, $Linear$ and $Uses$-$\beta$ ($\beta$ is fixed to $0.5$), we generate a total of 30 000 stego bags plus 5 000 cover bags, with inside each bag a bit-rate of $\bar{R}$ = 0.1 bptc. Whatever the pooling function, a set of 5 000 pairs of cover/stego bags (5 000 cover bags and 5 000 stego bags) is used. For each bag, the Rich Model, and the SID score is computed leading to a vector of SID scores. All the process is illustrated in Figure $\ref{fig:splitting_data}$.



\subsection{$g_{disc}$ learning}
\label{sec:g_disc-learning}

Given the 35 000 bags obtained for a bag size $b \in \mathcal{B}$ and all the spreading strategies ($IMS$, $DeLS$, $DiLS$, $Greedy$, $Linear$ and $Uses$-$\beta$), the minimum and the maximum of the SID scores are computed, which allows to define $p = 100$ centers, equally spaced in the range [$min_{x\in\mathcal{X}} f(x), max_{x\in\mathcal{X}} f(x)$], as in \cite{pevny2015optimizing}. A Parzen histogram {\bf h} can then be computed for each of the 35 000 bags. The pooling function $g_{disc}$, uses for the learning, 5 000 cover bags (i.e 5 000 Parzen vectors ${\bf h}$) and their corresponding 5 000 stego bags (i.e 5 000 Parzen vectors ${\bf h}$) equally distributed on the six spreading strategies. Note that $g_{clair}$ uses 5 000 cover bags and their corresponding 5 000 stego bags for one strategy.


The classifier, used for learning $g_{disc}$, is a SVM with a linear kernel. We use the SVM package from the free software machine learning library for the Python programming language Scikit-Learn \cite{scikit-learn}. The parameters are set to default, but the value of the kernel is set to 'linear'.

\section{Results}
\label{sec:Results}



As shown in Figure \ref{fig:splitting_data}, tests are done by using 2 500 cover image never seen, which allow to form a set of 5 000 pairs cover/stego bags for a fixed size $b$ for an spreading strategy ($IMS$, $DeLS$, $DiLS$, $Greedy$, $Linear$ and $Uses$-$\beta$). In order to generate a stego bag, a set of $b$ covers is randomly picked among the 2500 images, and the spreading strategy is then executed. The average probability of error, obtained by each pooling function, over 10 runs done each time with a different learning set (and testing set) of 5000 pairs of cover/stego bags, is then reported. 

\begin{figure}
\vspace{-0.2cm}
\includegraphics[width=0.48\textwidth, height= 4cm]{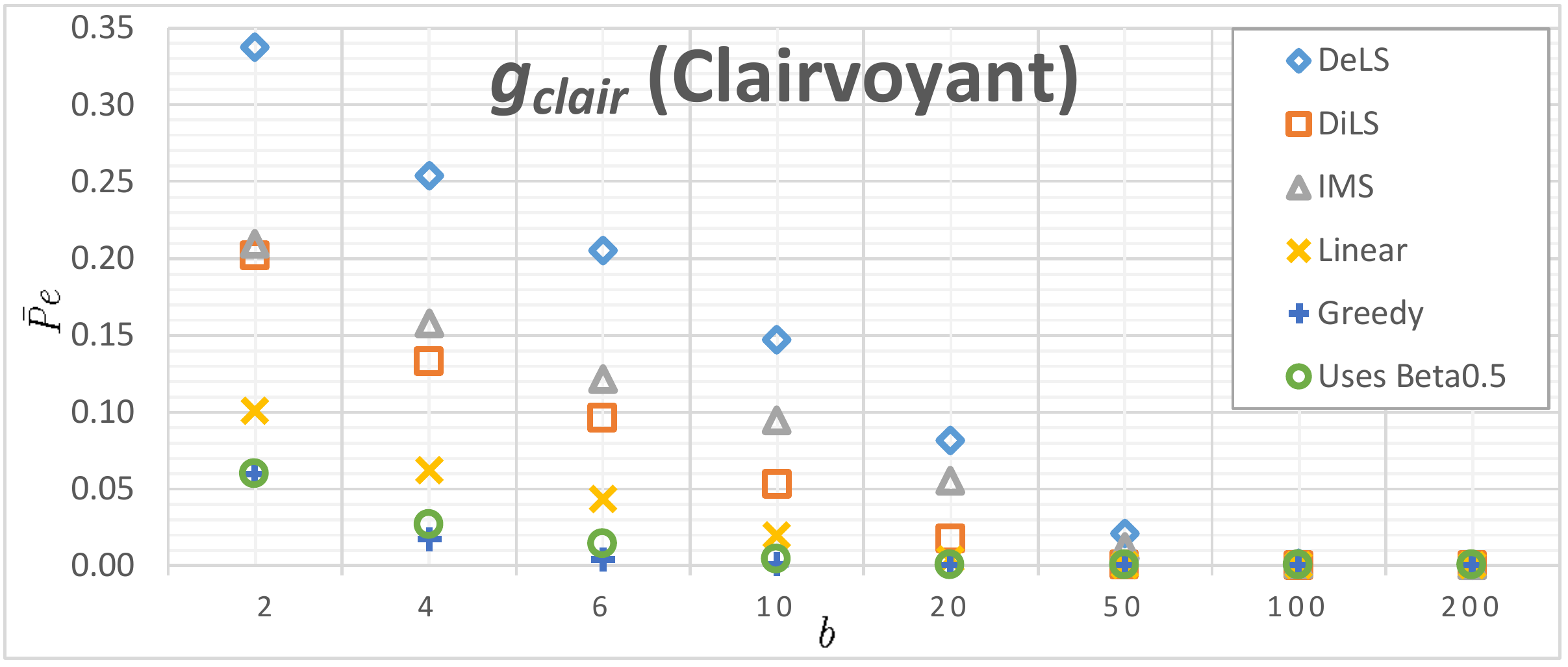}
\vspace{-0.1cm}
\caption{Spreading strategies comparison in the {\it clairvoyant} case. Average probability of error under equal prior, $\overline{Pe}$, as a function of pooling bag size $b \in \mathcal{B}$ for an average payload 0.1 bptc for $g_{clair}$ pooling function.}
\label{fig:g_disc_clairvoyant_pe}
\vspace{-0.4cm}
\end{figure}


Figure \ref{fig:g_disc_clairvoyant_pe} reports the results obtained with the {\it clairvoyant} steganalysis scenario i.e. with $g_{clair}$. One can notice that the DeLS is the best and it outperforms the $IMS$, which was more competitive to $DeLS$ in \cite{cogranne2017practical}, while the Greedy strategy is the worst. 
One can cluster the strategies into 3 groups: the $Greedy$ and $Uses$-$\beta$ which are highly detectable with a detectability which starts to coincide for bag sizes $\ge$ 10 with $\overline{Pe}\approx0$, the strategies $DeLS$, $IMS$ and $DiLS$ which are the more secure ones, and the $Linear$ strategy which falls between the two other groups. The strategies $DeLS$ and $IMS$ becomes totally detectable at $b=100$ with $\overline{Pe} \approx0$.

\begin{figure}
\vspace{-0.2cm}
\includegraphics[width=0.48\textwidth, height= 4cm]{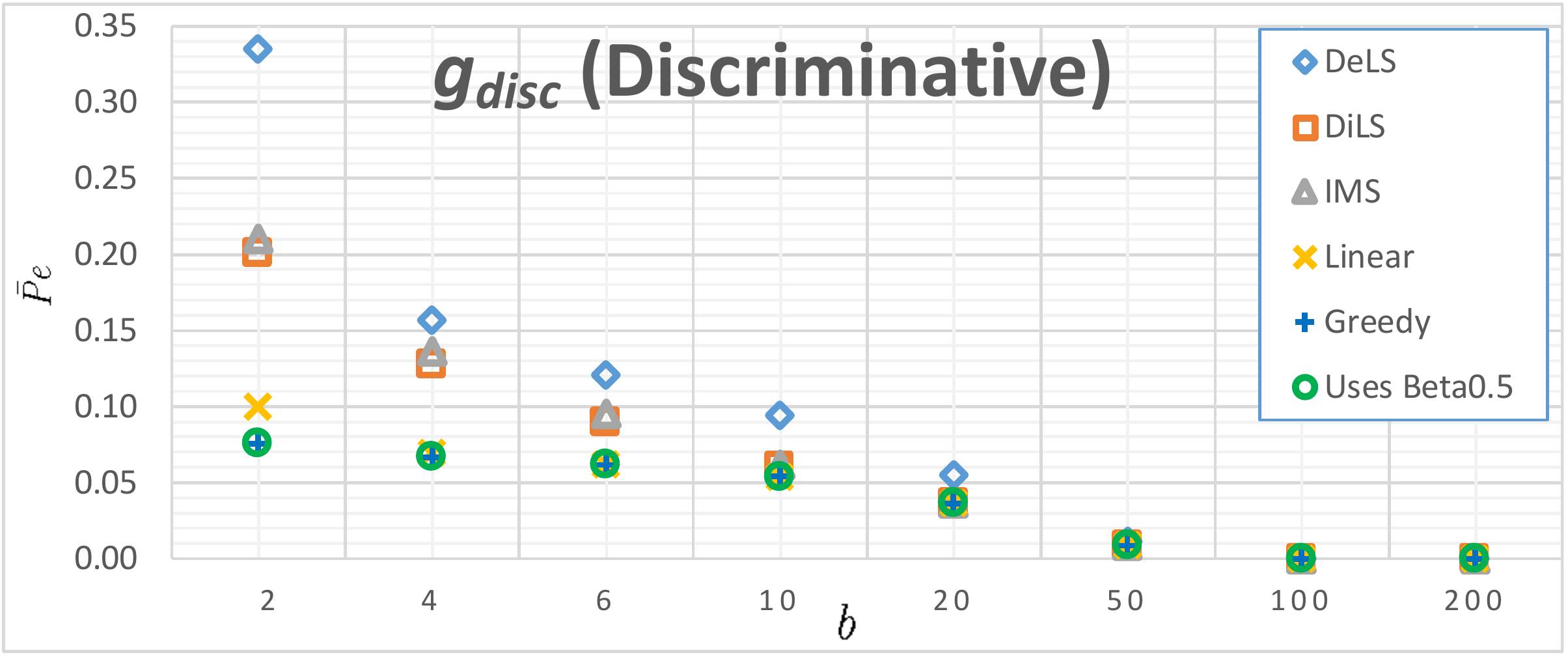}
\vspace{-0.1cm}
\caption{Spreading strategies comparison in the {\it discriminative} case. Average probability of error under equal prior, $\overline{Pe}$, as a function of pooling bag size $b \in \mathcal{B}$ for an average payload 0.1 bptc for $g_{disc}$ pooling function learnt over all the strategies.}
\label{fig:g_disc_pe}
\end{figure}

In Figure \ref{fig:g_disc_pe}, we report the detection of each spreading strategy with the {\it discriminative} $g_{disc}$ pooling function. The best strategies are in the descending order, in the sense of its ability to resist the pooling function that tries to {\it discriminate} it, $DeLS$, $IMS$, $DiLS$, $Linear$, $Uses$-$\beta$, $Greedy$. $DeLS$ is again performing well in this {\it non-clairvoyant} approach, and it remains resistant until $b=100$ where its average probability of error $\overline{Pe}$ start to coincide with those of the other strategies and becomes $\approx0$. 
$DeLS$ and $IMS$ are more detected by the discriminative $g_{disc}$ than the clairvoyant $g_{clair}$. Looking at the histograms of the SVM scores, and to the ROC curves (not shown in this paper), we observe indeed a higher separation with $g_{disc}$. Looking to Figure~\ref{fig:g_disc_pe}, we also observe a smaller gap between all the $\overline{Pe}$ of each strategy, compared to Figure~\ref{fig:g_disc_clairvoyant_pe}. Those behaviour are probably because the optimization (learning of the SVM) try to minimize the prediction error fairly for each of the strategies.

\begin{figure}
\vspace{-0.2cm}
\includegraphics[width=0.48\textwidth]{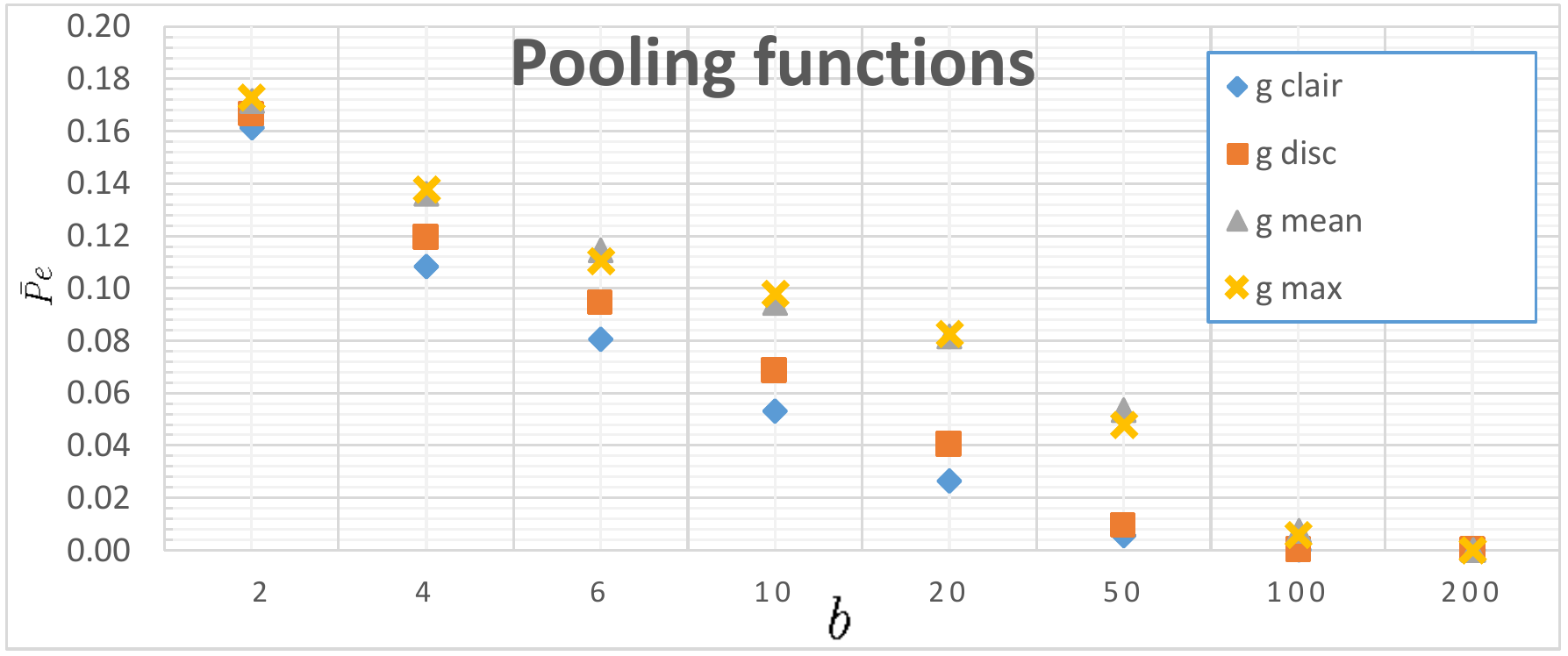}
\vspace{-0.1cm}
\caption{Pooling steganalaysis comparison. Average probability of error under equal prior, $\overline{Pe}$, as a function of pooling bag size $b \in \mathcal{B}$ for an average payload 0.1 bptc. The average $\overline{Pe}$ is computed by testing each spreading strategy.}
\label{fig:avg_strategy_Pe}
\vspace{-0.4cm}
\end{figure}

Finally, Figure \ref{fig:avg_strategy_Pe} provides a useful insight regarding the accuracy of the pooling methods. This figure shows the {\it average} probability of error under equal prior, $\overline{Pe}$, as a function of pooling bag size $b \in \mathcal{B}$, with an average payload size $\bar{R}$ = 0.1 bptc for each pooling function over all the strategies. We note that the $g_{disc}$ pooling function outperforms $g_{mean}$ and $g_{max}$ functions with average difference of $\overline{Pe}\approx 2\%$  
and is closer to $g_{clair}$ with an average difference of $\overline{Pe}\approx0.8\%$\footnote{The variance on $\overline{Pe}$ for each pooling is around $\times{10^{-6}}$.}. 
From figure \ref{fig:avg_strategy_Pe}, we could observe that $g_{disc}$ and $g_{clair}$ are more stable than $g_{mean}$ and $g_{max}$. This is in agreement with the hypothesis of our paper which is that a {\it discriminative} pooling function allows obtaining better detection results compared to the mean and max pooling function.


\section{Conclusion}
\label{sec:conclusion_and_perspectives}
In this paper, we studied the problem of content-adaptive batch steganography and pooled steganalysis for a steganalyst unaware of the payload-spreading strategy and equipped with a single-image detector trained as a quantitative classifier.

We studied the ability of the steganalyst to {\it discriminate} the spreading strategy, thanks to a pooling function that is able to recognize various stego patterns, and then able to pool the SID scores much cleverly than applying a simple average or maximum.  Empirical results made with six different spreading strategies and a state-of-the-art Single Image Detector confirms that our {\it discriminative} pooling function can improve the accuracy of the pooled steganalysis. Our pooling function gets results close to a $clairvoyant$ steganalyst which is assumed to know the spreading strategy.

This conclusion opens the door to future studies related to performance of a steganalyst that would not be aware of the bag's payload, to the performance of the steganalyst in the case of a spreading strategy never seen before, to the performance of an all integrated solution using Deep Learning, to the extension to a game-theory or practical GAN simulation, etc.

\vspace{-0.36cm}
\section*{Acknowledgment}
\vspace{-0.1cm}
{\footnotesize We would like to thank Dr R\'emi Cogranne for discussions and for providing the MATLAB code of the DiLS, DeLS and IMS. We would like to thank the MESO@LR 
computing center, Montpellier, France, for providing a huge amount of calculation resources. We would like to thank the French Direction G\'en\'erale de l'Armement (DGA) for its support through the Alaska project ANR (ANR-18-ASTR-0009). We would like to thank the Union of Municipalities of Jered Al-Kaytee, Akkar, Lebanon, for its scholarship support.}
    


\vspace{-0.05cm}
\bibliographystyle{IEEEtran}
\vspace{-0.2cm}
\bibliography{biblio}
%

\end{document}